\documentclass[conference]{IEEEtran}

\usepackage{amsthm,amssymb,graphicx,multirow,amsmath,color,amsfonts}
\usepackage[update,prepend]{epstopdf}
\usepackage[noadjust]{cite}
\usepackage[latin1]{inputenc}
\usepackage{bbm} 
\usepackage{pdfpages}
\usepackage{tabulary}
\usepackage{multirow}
\usepackage{comment}
\usepackage{array}
\usepackage{tabularx}

\usepackage{amsmath,amssymb}

\IEEEoverridecommandlockouts

\let\svbibcite\bibcite
\def\bibcite#1#2{\svbibcite{#1}{#2}}
\makeatletter
\let\svbiblabel\@biblabel
\def\@biblabel#1{\svbiblabel{#1}}
\makeatother
\usepackage{fancyhdr}
\renewcommand{\thispagestyle}[1]{} 

\usepackage{tikz}
\usepackage{textcomp}
\usepackage{lipsum}

\newcommand\copyrighttext{  
 This work has been submitted to the IEEE for possible publication. 
 Copyright may be transferred without notice, after which this version 
 may no longer be accessible.
 
	}
\newcommand\copyrightnotice{%
	\begin{tikzpicture}[remember picture,overlay]
	\node[anchor=north,yshift=-7pt] at (current page.north) {\fbox{\parbox{\dimexpr\textwidth-\fboxsep-\fboxrule\relax}{\copyrighttext}}};
	\end{tikzpicture}%
}

\begin{document}  

\title{  Towards 6G with Connected Sky:\\ UAVs and Beyond 	}  

\author{\IEEEauthorblockN{ Mohammad Mozaffari, Xingqin Lin, and Stephen Hayes}\vspace{-0.25cm}\\
	\IEEEauthorblockA{
	\small Ericsson\\
	\small Emails:{\{mohammad.mozaffari, xingqin.lin, stephen.hayes\}@ericsson.com}
\vspace{-0.2cm}}
		
	}

\maketitle 
 \copyrightnotice 
\begin{abstract}
	
	The large-scale and ever-growing use of unmanned aerial vehicles (UAVs) in a wide range of applications is foreseen to be a major part of beyond 5G and 6G wireless networks in the next decade.  The effective support of such massive deployment of UAVs requires offering reliable, secure, and cost-effective wireless connectivity. In this regard, cellular networks play essential roles in serving UAVs acting as flying user equipments. While the cellular networks provide
	promising connectivity solutions for UAVs, enabling robust UAV operations faces several challenges. In this paper, an overview of key barriers and design considerations of widespread commercial use of flying UAVs are presented along with their potential solutions. In addition, we discuss how cellular networks can support UAVs by relying on their advanced features, network intelligence, \textcolor{black}{key enabling technologies for beyond 5G and 6G,} and exploiting new tools from machine learning. Finally, we shed light on offering wireless services to high altitudes and the integration of non-terrestrial networks with terrestrial networks towards limitless connectivity in 6G.  

\end{abstract}

\section{Introduction}
The next generation wireless network towards 6G is envisioned to enable intelligent, secure, reliable, and limitless connectivity \cite{saad2019vision,E6G}. It is expected that 6G will bring a full-fledged framework for connected things and automation systems from autonomous cars to unmanned aerial vehicles (UAVs) with stringent and diverse requirements in terms of reliability, latency, data rate, and energy efficiency. UAVs, commonly known as drones, play key roles in a wide range of use cases and scenarios which can go beyond 5G and 6G \cite{saad2019vision, MozaffariB5G}. 
 The examples of UAV applications include package delivery, media production, real-time surveillance, and remote constructions.  
 The deployment of  UAVs is ever-increasing and, as predicted by the Federal Aviation Administration (FAA), the number of commercial UAV fleets  can reach up to 1.6 millions by 2024 \cite{FAAFiscal}. 
   Meanwhile, unmanned aircraft systems (UAS) traffic management (UTM) systems are being developed to support safe and secure operation of low-altitude UAVs with their potentially high traffic density.

\begin{figure}[!t]
	\begin{center}
		\includegraphics[width=8cm]{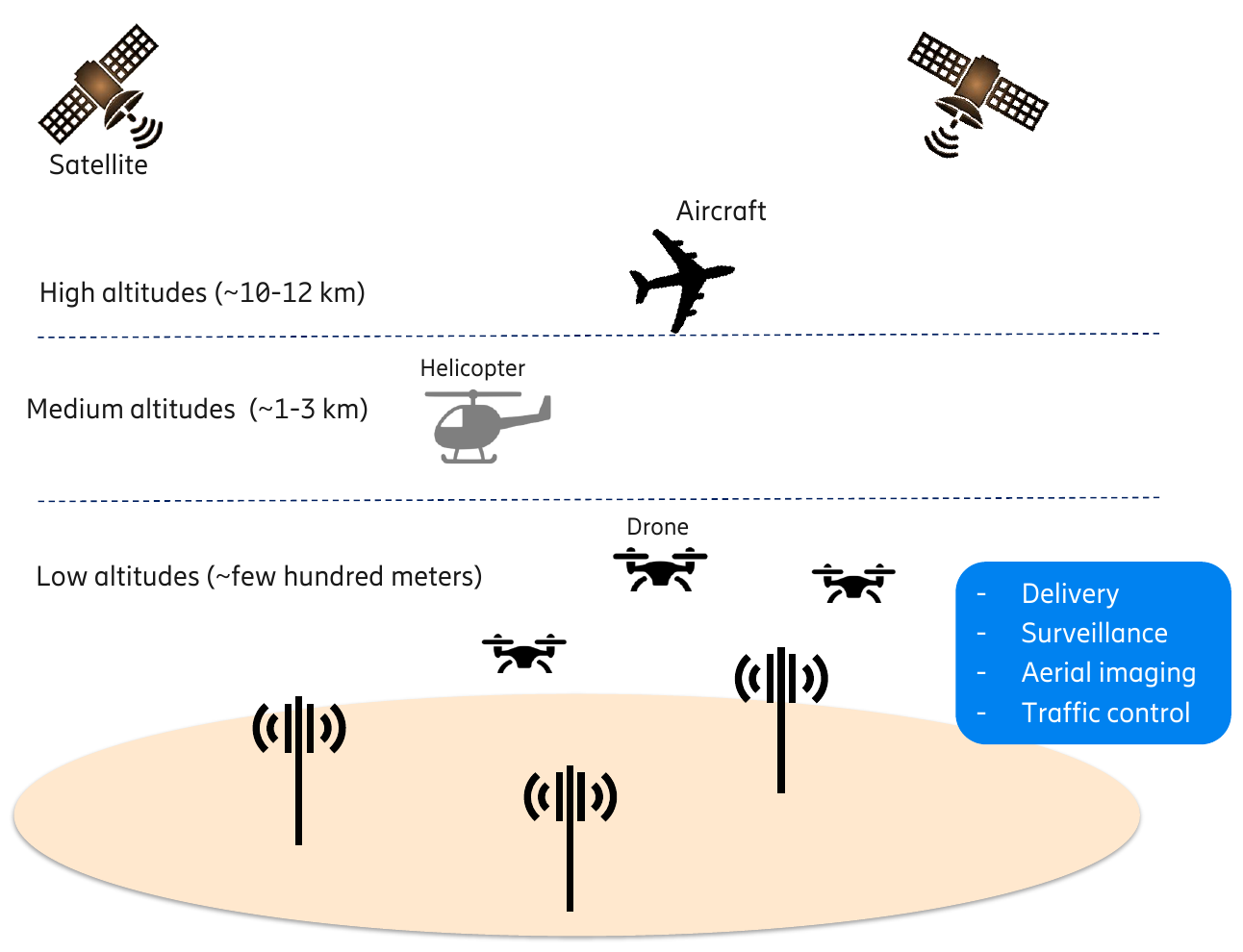}
		\vspace{-0.2cm}
		\caption{Cellular-connected UAVs and manned aircraft with different altitudes.}\vspace{-0.6cm}
		\label{Fig1}
	\end{center}
\end{figure}

 In a connected sky, the support of large-scale deployment of UAVs requires reliable and secure wireless communications that ensure safe control and  operation of UAVs. This will require efficient system design for wireless communication, intelligent computation, and reliable control mechanisms. Specifically, to support UAV operations in various applications, the following aspects can be realized: i) reliable three-dimensional (3D) wireless connectivity with enhanced cellular networks and integration of terrestrial and non-terrestrial networks (NTNs), ii) artificial intelligence (AI) at the edge used for UAV image processing in scenarios such as search and rescue, collision avoidance, and inspections, and iii) robust control and inter-UAV communications for flying in formation and cooperative operations. {\color{black} The terrestrial cellular networks, potentially in integration with NTNs,  can provide robust, reliable, and limitless wireless connectivity for UAVs  flying at different altitudes (as illustrated in Fig.\,\ref{Fig1}).}   

The cellular systems  have the ability to serve flying UAVs by providing wide-area, cost-effective, and reliable wireless connectivity. However, the existing cellular infrastructure is primarily optimized for serving user equipments (UEs) located on (or close to) the ground. In this regard, the efficient support of aerial UEs, particularly in the next generation wireless networks which need to provide limitless connectivity, faces new challenges due to their relatively high altitude, mobility, massive deployment, and flight safety requirements. Subsequently, the design of cellular-connected UAV systems for supporting flying UEs has received significant attention in both academia and industry \cite{mozaffari2019tutorial, lin2019mobile, TR36.777}.  The studies on cellular-connected UAVs focus on various areas including standardization, mobility support, deployment optimization, and performance evaluations \cite{mozaffari2019tutorial, saad2020wireless}.  Meanwhile, the 3rd generation partnership project (3GPP) started standardization activities for supporting aerial UEs in Release-15 followed by further work in the subsequent 3GPP releases. 

Compared to the traditional cellular-connected terrestrial UEs, cellular-connected UAV-UEs possess different characteristics that result in several tradeoffs. For example, compared to the ground UEs, UAV-UEs experience more line-of-sight (LoS) channels due to their higher altitudes. Such LoS links between UAVs and base stations (BSs) are beneficial as they reduce the signal blockage and lead to high reference signal received power (RSRP). {\color{black} Although such LoS condition is beneficial for the serving BS,  interfering BSs can also generate strong LoS interference which degrades the communication link quality.  Hence, the cellular-connected UAVs suffer from strong interference in downlink due to nearly LoS channel conditions between BSs and UAVs.  Similarly, the uplink interference can be severe when the number of transmitting UAVs is large. This, in turn, makes it challenging to provide reliable connectivity in the sky.} Further, the UAV mobility support and handover management in the sky are challenging due to the relatively high speed of UAVs and BS antenna patterns \cite{chen2020efficient}. 

In this paper, we provide an overview of key barriers and technical challenges associated with a widespread commercial use of UAVs towards 6G and propose a number of promising solutions. We first focus on the foundation of safety, traffic control, and security aspects needed for supporting UAVs. Then we discuss how the cellular networks can effectively serve UAVs by relying on their advanced features and exploiting new solutions from AI. Furthermore, we look into beyond low-altitude coverage for offering broadband in-flight connectivity to aircraft that fly at high altitudes. Finally, we discuss the opportunity of complementing terrestrial networks with NTNs which can be a component of 6G for providing limitless wireless connectivity.


\section{Key Barriers of Widespread Commercial Usage of UAVs and Solutions}

In this section, we examine three key barriers of the wide-scale UAV deployment and how the cellular technology can mitigate them.  

\subsection{Safety of UAVs}
The aviation community has an admirable flight safety record.  It is understandable that they wish to ensure that the UAVs are operated safely as they become more and more widespread.  A key requirement for aircraft has always been that the pilot is in control of the aircraft at all times.  When the pilot is not onboard, then the reliability of the communications link between the pilot and the UAV becomes critical. 

In principle, the required reliability level can depend on methodologies such as SORA (Specific Operations Risk Assessment).  Proving that the network and UAV have the necessary reliability and availability at the planned altitudes over the operational area is a daunting task. Safe operation must also consider the need to operate in challenging or adversarial environments where UAVs may be faced with attacks such as interference, spoofing, or jamming. \textcolor{black}{Indeed, reliability is a crucial  metric for UAV operations which can be represented in terms of mean time between failures. The reliability (i.e., failure rate) requirement for commercial aviation is around one per $10^5$ flight hours.  For drone systems, the reliability  is approximately one per $10^3$  flight hours, which is 100 times lower than the commercial aviation failure rate.} 

To address this challenge GUTMA (Global UTM Association) and GSMA (Global System for Mobile communications Association)  are developing a long-term evolution (LTE) aerial profile.  This profile specifies the required service levels and capabilities in LTE networks and devices to operate drones reliably.  Additional work is ongoing on developing minimum performance specifications for modules in UAVs and testing methodologies and is planned for 5G new radio (NR) once the necessary performance improvements are identified by 3GPP.

3GPP systems provide multiple quality of service (QoS) classes to fit the needs. However, having various QoS requirements such as high reliability as well as high throughput comes at a price.  Many UAS manufacturers have therefore taken the approach of introducing onboard intelligence in their drones to allow them to avoid collisions and perform basic operations while satisfying QoS requirements.  This both reduces the granularity of the communications  as well as introducing the ability to survive brief communications outages.  Many UAS manufactures introduce multiple technologies such as cellular or satellite to provide redundancy.

\subsection{Security of UAVs}
There have been multiple high profile incidents in which UAVs have been in areas where they should not be, sometimes accidentally and sometimes for nefarious goals such as assassination attempts.  A key step in enforcing proper use of UAVs is to be able to identify drones.  Like larger aircraft, UAVs must be usually registered. For example, in the US, all drones with a weight greater than 0.55 pounds must be registered.  That registration ID must be on the body of the UAV.  However, the registration sticker may be quite small and would be  hard to read from the ground.  For this reason, the FAA in 2021 has required that all UAVs ($> 0.55$ pounds) support Remote ID. The form of Remote ID currently required is a periodic broadcast beacon transmitted over WiFi or Bluetooth.  These can be received by an appropriately configured smartphone.   This solution has limitations since the broadcast is on unlicensed bands, it is not clear that it will be sufficiently reliable in congested urban environments.  It also requires that a person be present near the UAV with an appropriate device to receive the broadcast.

To overcome these limitations, a second form of Remote ID was developed.  This is mobile network-based remote ID where the UAV periodically transmits its location, vector, status, etc. to a remote database.  That database can then be queried to find what UAVs are in an area and associated information such as who owns the drone and its mission.  That solution, however, relies on having a federation of databases to handle this and other related queries.  This federation does not exist yet and that is one of the reasons for deferring the second remote identification method.  In this regard, 5G and beyond are the prime candidates to efficiently transfer such tracking information to a remote database.

\subsection {UAVs and airspace control}
In terms of airspace control, there are extensive procedures governing their operations as well as flight plans and tight orchestration of their position within the airspace.  The aviation community is understandably concerned with having large numbers of UAVs flying around without robust tracking of where they are.  However, due to the sheer number of UAVs, this process needs to be automated.  The system handling drone air traffic management is referred to as the UTM. It is worth noting that the UTM is not a single system, rather is envisioned as a federation of systems that authorize and track UAV traffic. The UTM provides a variety of flight related functions for UAVs and UAV operators.  For example, the network based remote ID mentioned earlier would periodically publish location information to the UTM. 

\begin{figure}[!t]
	\begin{center}
		\includegraphics[width=8.3 cm]{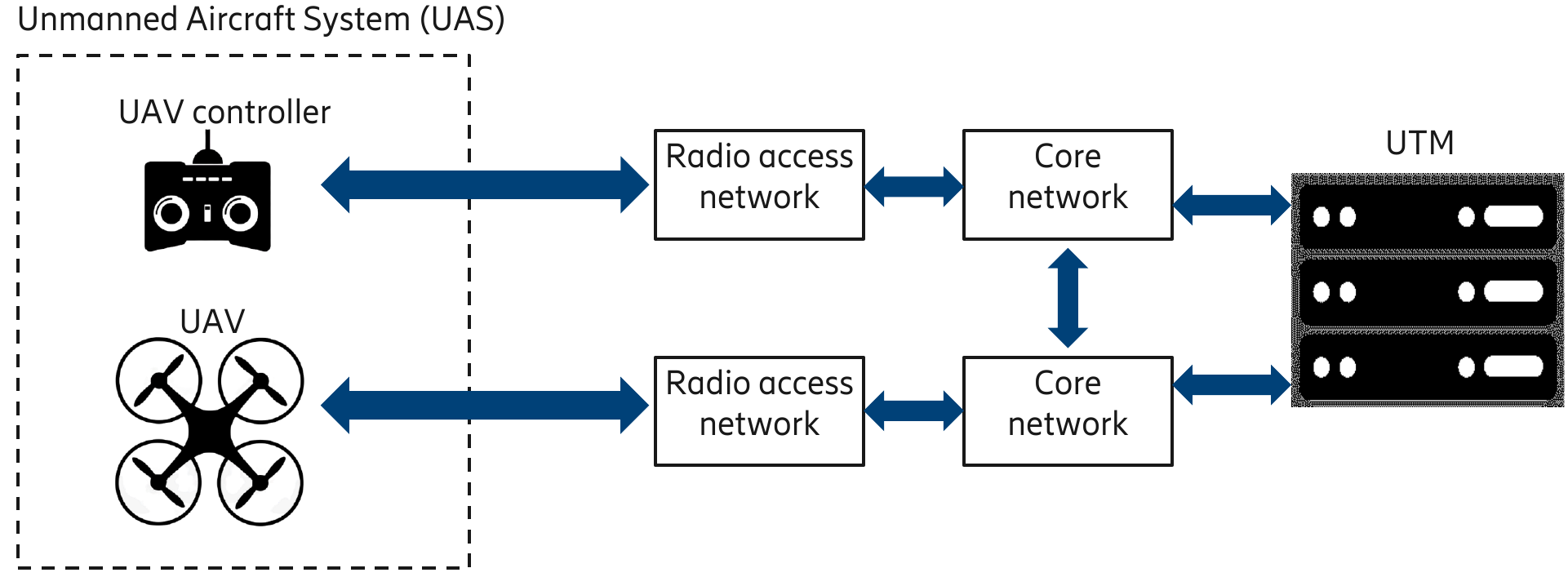}
		
		\caption{Illustration of the UAS-to-UTM connectivity.}\vspace{-0.4cm}
		\label{Fig3}
	\end{center}
\end{figure}

 A UAS is composed of a UAV and a UAV controller used by an operator with unique credentials and identities. Mobile networks can enable reliable connectivity between the UAV and its controller and UTM can connect to the UAV and the UAV controller through the core network and the radio access network, as illustrated in Fig.\,\ref{Fig3}.

When the UAV is using LTE or 5G networks for connectivity, additional benefits can be gained.  Work is ongoing in 3GPP to integrate UTM communications with the network and thus provide additional capabilities.  Examples of the types of additional capabilities that can be provided are:
\begin{itemize}
\item Linking flight authorization and communications resource allocation.
\item 	Integrity checks on the flight plan using network-based positioning information.
\item 	Mapping between various identifiers.
\end{itemize}

 \section{Enhanced Cellular Network Capabilities for UAVs}
 
In this section, we discuss the key differences between flying UEs and terrestrial UEs, describe existing features of cellular networks for supporting UAVs, and identify novel solutions from AI that boost UAV connectivity in the sky.
 
\subsection {UAV-UEs versus terrestrial UEs}\label{Challnge}
\begin{figure}[!t]
 	\begin{center}
 		\includegraphics[width=8.8cm]{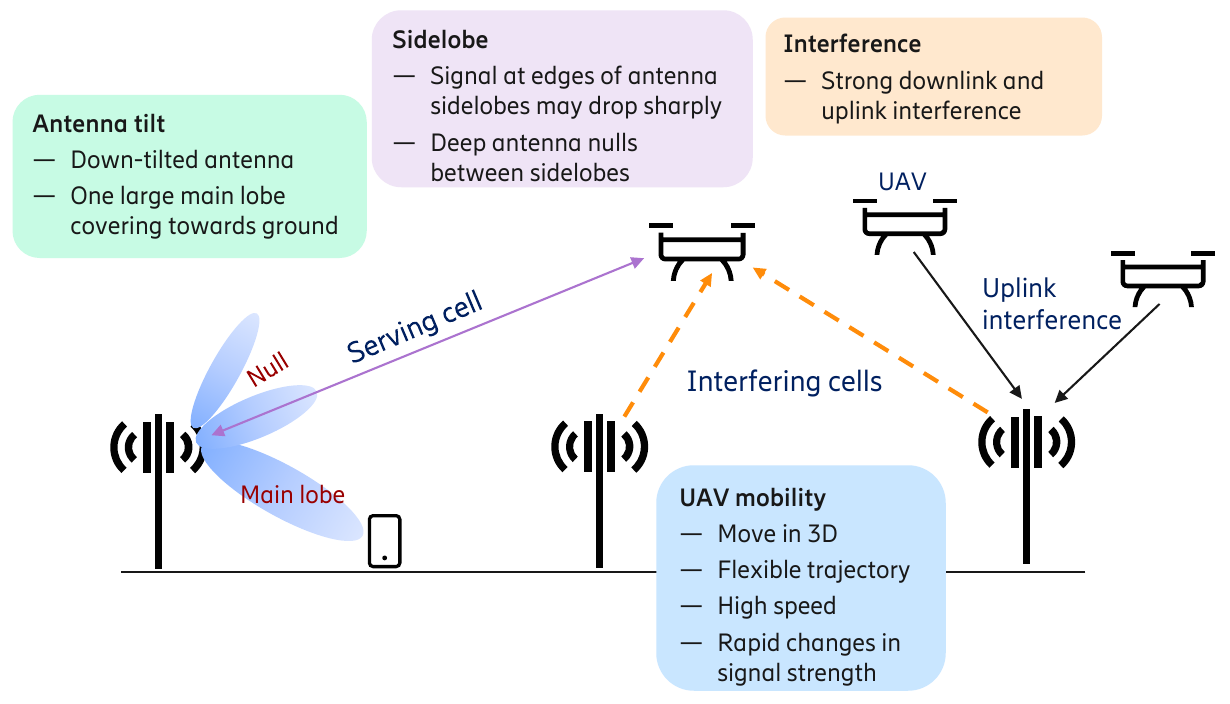}
 		\vspace{-0.5cm}
 		\caption{Challenges of cellular-connected UAV.}\vspace{-0.4cm}
 		\label{Fig2}
 	\end{center}
 \end{figure}

 Compared with terrestrial UEs, efficient connectivity support for flying UAVs requires coping with several challenges,  as summarized below and illustrated in Fig.\,\ref{Fig2}.
 
{\color{black} \textbf{3D mobility:} UAVs can move in a 3D space and potentially with high speeds. Such 3D mobility that involves considerable change in the UAV's altitude  significantly impacts the propagation channel characteristics between UAVs and ground BSs. This, in turn, can result in a rapid fluctuation of the received signal power and link quality of cellular-connected UAVs.}  
 
 \textbf{{\color{black}BS antenna tilt and sidelobes:}} Terrestrial BSs are designed for serving UEs which are close to the ground. Therefore, the main lobes of BS's antennas face towards ground and UAVs flying at relatively high altitudes are typically served by one or multiple sidelobes with smaller beamwidth and antenna gain. In addition, in the BS's antenna pattern, there exists several nulls that prevent the BS from providing continuous coverage in the sky (i.e., having coverage holes in the sky). Due to such BS's antenna pattern and special positions of UAVs with respect to BSs, the cell association pattern becomes fragmented in a multi-BS network. This issue can be mitigated by employing efficient adaptive 3D beamforming solutions to offer coverage to the sky \cite{mozaffari2019tutorial}. 
 
\textbf{Strong interference:} Given the LoS condition in BS-to-UAV communications, UAVs can experience severe downlink interference from neighboring cells. Similarly, in uplink communications, the dominant LoS channel between UAVs and BSs may result in strong interference. 
 
 Despite the challenges of providing connectivity for UAV-UEs, they have several advantages over regular ground UEs which are summarized below.
 
 \textbf{Predictable path:}  UAVs' routes are typically predictable based on their missions. Such predictable paths allow creating sky corridors to efficiently track and serve UAVs. 
 
 \textbf{Multi-subscriber identity module (SIM):} {\color{black} Multi-SIM allows a device to have service from multiple mobile network operators. This is particularly beneficial for  UAVs flying in the sky that may need to switch to a different operator network when the current operator network cannot provide the required coverage.   The multi-SIM technology  facilitates seamless UAV connectivity while enhancing the coverage, security, and safety of UAV operations.}
 
 
 \textbf{Controlled density:} UAVs are typically density controlled since UTM enforces specific flight separations. Such controlled density is beneficial for traffic control, interference 
 management, and network optimization. \vspace{-0.01cm}

 \subsection{Cellular-connected UAVs: From LTE to 5G NR}
 
 In 3GPP, the study on the cellular-connected UAV initiated in Release-15 in order to evaluate the capability of the LTE network in supporting UAV-UEs. This study concluded that LTE networks can properly serve flying UAV-UEs, as long as challenges pertaining to the downlink and uplink interference, the mobility support in the sky, and potential impacts on the terrestrial UEs are addressed. To tackle these issues, several implementation-based and specification-based solutions were identified. Subsequently, a work item in Release-15 was conducted to specify the features that can enhance the efficiency and reliability of LTE-connected UAV networks \cite{TS36.331}. 
 In addition, the 3GPP Release-16 study item on remote UAV identification explored the potential requirements and use cases for remote identification and also identified services that can be offered based on the remote identification of UAVs \cite{TR22.825}.
 
 The cellular-connected UAV network is further evolving towards the 5G NR-enabled UAV. In addition to potentially adopting features and techniques introduced in LTE and 5G NR has advanced capabilities that allow robust support of flying UAVs.  Specifically, the 5G NR capabilities include flexible numerology (e.g., subcarrier spacing), more dynamic scheduling, advanced antenna technologies and beamforming capabilities, more sophisticated carrier aggregation scheme, as well as wide bandwidth operations. These features and capabilities of 5G NR are the key enablers of providing superior performance in terms of 3D coverage, data rate, latency, and reliability for supporting UAV connectivity. Meanwhile, the 3GPP Release-17 studied 5G enhancement for UAVs \cite{TS22.125}. In this work, new key performance indicators (KPIs) and communication requirements of UAVs with a 3GPP subscription are identified, which include KPIs for communication service and for command and control traffic. Also, the 3GPP Release-17 is studying application layer support for UAS, including functionalities to support and interact with UTM \cite{TR23.755}.

{\color{black}The 5G capabilities and its advanced features enable efficient support of UAV communications.  More specifically: 
 
 \begin{itemize}
 \item Spectrum flexibility: 5G supports flexible numerology, large bandwidth up to 400 MHz in high frequency bands (above 24 GHz), and carrier aggregation with up to 16 component carriers, which enables high data rate communications. For  UAVs which are used in virtual reality (VR) applications, such high data rate wireless communication is required for transmitting high-quality images and videos. In addition, 5G supports operation in the spectrum ranging from sub-1 GHz to millimeter wave. Combining the uses of low-band, mid-band, and high-band spectrum in 5G deployments achieves highest quality performance for UAV communications.
 
 \item Advanced antenna technologies: 5G relies on advanced antenna technologies such as enhanced beamforming and massive MIMO schemes while supporting single-user and multi-user MIMO. With eight and four transmission layers for single-user MIMO in downlink and uplink, respectively, significant capacity increase can be achieved in cellular-connected UAV systems. Furthermore, 5G exploits beam management and beam-sweeping features with narrow beam operations that can boost coverage and capacity while reducing interference for  supporting UAVs in the sky.

 \item Forward compatibility:  A unique aspect of 5G NR is forward compatibility that allows introducing new features and components for supporting emerging use cases. To efficiently support the widespread use of UAVs in new use cases over time, there will be a need for introducing new compatible features in the future.    
 
 \item Network slicing: This enables multiple logical networks to efficiently operate using a shared physical infrastructure while providing different services. It allows network optimization for meeting  various requirements (e.g., latency, security, and  reliability) of  UAV command and control, and data transmissions in different applications. 
 
 \item Ultra-lean design: The goal of ultra-lean design is to  minimize \emph{always on} transmissions particularly for reference signals and synchronization signals.  Ultra-lean design is beneficial for energy efficiency and interference mitigation which are key considerations in cellular-connected UAV networks.
 \end{itemize}
}

 Moreover, network intelligence, which relies on AI and big data analytics, can be employed for extracting information from existing data as well as modeling and optimizing complex cellular-connected UAV networks. One important example of AI-assisted UAV communications is UAV mobility support in the sky, which is described in the following section. \vspace{-0.3cm}
 
 
 \subsection{Embracing artificial intelligence}
 {\color{black} The performance of cellular-connected UAV networks can be further enhanced by leveraging tools from  AI and in particular machine learning (ML) techniques.} Network intelligence which is envisioned to be a key feature of 6G can play an essential role in UAV connectivity. Specifically, AI-based solutions can be used for efficient UAV identification, traffic management, UAV flight path planning, and collision avoidance. In addition, ML can assist in network deployment optimization, resource management, handover optimization, and mobility support in the sky. 
 
 Efficient UAV mobility support is essential for coping with challenges of cellular-connected UAV networks described in Section \ref{Challnge}. For example, the scattered RSRP map for UAVs can result in frequent handovers, radio link failure, and ping-pong handover events. Therefore, providing seamless UAV connectivity in the sky requires employing robust handover mechanisms. In this regard, reinforcement learning (RL) can play a key role in developing optimal handover rules that ensure reliable connectivity with a minimum handover cost.
 
\begin{figure}[!t]
 	\begin{center}
 		\includegraphics[width=8 cm]{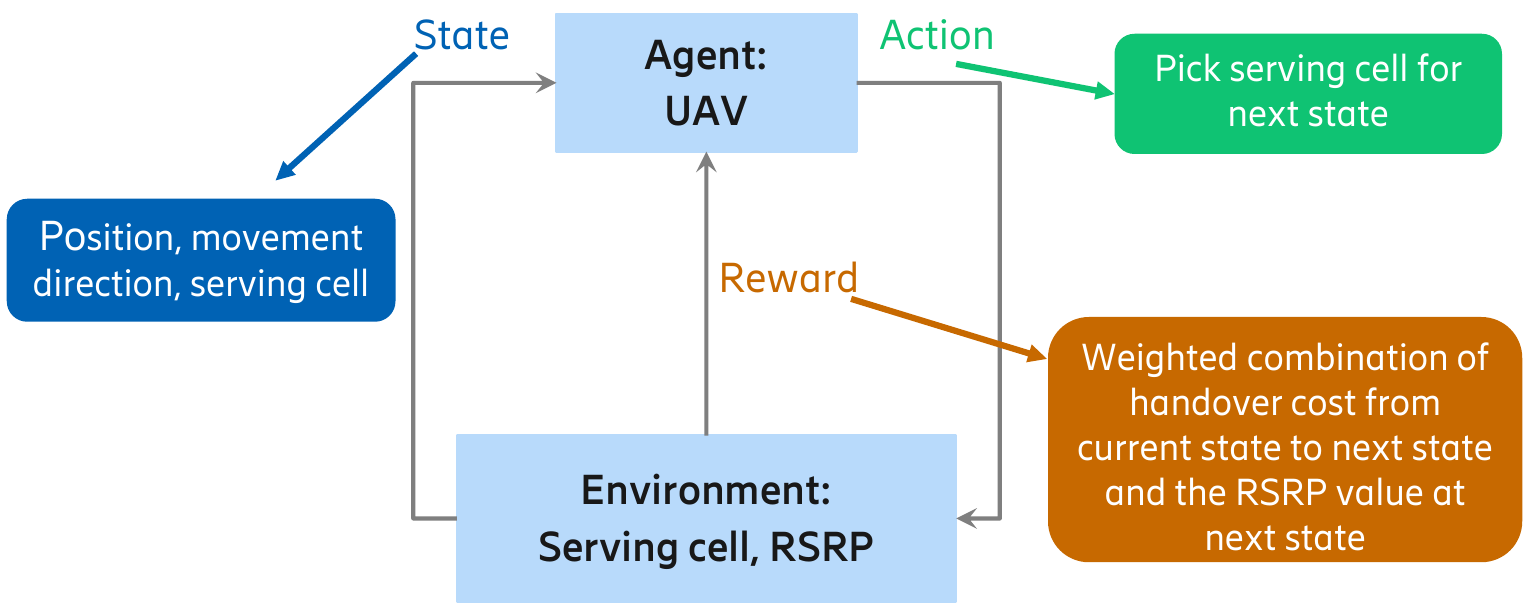}
 		\vspace{-0.2cm}
 		\caption{An RL-based handover mechanism.}\vspace{-0.5cm}
 		\label{Fig4}
 	\end{center}
 \end{figure}

 As a representative result, we compare the performance of an RL-based handover scheme \cite{chen2020efficient} with a greedy approach in which a UAV always connects to the strongest BS (i.e., with highest RSRP value). Given the fragmented cell association pattern in the sky, the greedy scheme results in a significant number of handovers.  The RL-based scheme, however, can capture the tradeoff between RSRP and the number of handovers using a reward function based on the weighted combination of RSRP values and the handover cost. In the RL scheme, an agent (e.g., {\color{black} UAV}) interacts with the environment, performs an action in a state, receives {\color{black}reward}/feedback based on its action, and moves to a new state. An illustration of the RL-based handover mechanism is shown in Fig.\,\ref{Fig4}.

The RL-based scheme can significantly reduce the number of handovers and the signaling overhead required for serving a flying UAV. For instance, for a UAV flying at a 50 m altitude over a geographical area with 21 cells (i.e., BSs), considering an equal weight of RSRP values and the handover cost, the RL-based scheme can yield around 10 times fewer handovers compared to the greedy approach \cite{chen2020efficient}.  \vspace{-0.01cm}

{\color{black}
\subsection{Enabling technologies of 5G and beyond}
In addition, cellular connected-UAV networks can exploit the key enabling technologies of 5G and beyond to enhance their performance. 

\textbf{Massive MIMO:}  By utilizing large antenna arrays, ground BSs can perform adaptive 3D beamforming with narrow beams to efficiently support UAVs in the sky and terrestrial users.  Massive MIMO can significantly enhance the spectral efficiency and coverage performance of cellular-connected UAV networks. 

\textbf{High frequency operation:} Millimeter-wave and terahertz (THz)  communications are promising solutions for offering high data rate services. Operating in such high frequencies offers wide bandwidth availability and allows using tiny antennas in massive MIMO systems which are key enablers for meeting high data rate requirements of UAV communications.   

\textbf{Intelligent surfaces:}  Reconfigurable intelligent surfaces (RISs) allow intelligently reflecting electromagnetic signals in a desired direction for improving propagation condition and communication link quality. Cellular-connected UAV systems can also exploit RISs to address the issue of down-tilted BS antennas by optimally reflecting transmitted signals towards UAVs in the sky and mitigate the downlink interference, which can lead to enhanced coverage for UAVs. 

\textbf{Energy harvesting and energy transfer:} With the advancement of technology towards 6G, energy harvesting and wireless energy transfer are becoming attractive for empowering battery limited devices. Similarly, for UAVs with battery constraints, energy harvesting (e.g., solar, wind) and wireless power transfer using ground BSs can be promising solutions for prolonging UAV operations.     	
}\vspace{-0.01cm}
 
 \section{Beyond Low-Altitude UAVs}
  
In the previous sections, we mainly focused on providing connectivity for low-altitude UAVs. Nonetheless, the need for limitless connectivity is ever-growing and  goes beyond low-altitude UAVs. Here, we discuss two aspects towards a further connected sky:   1) providing broadband connectivity to high-altitude aircraft using direct air-to-ground (A2G) communications, and 2) how NTNs can complement terrestrial networks.  \vspace{-0.6cm}
  
\subsection {Offering connectivity to high altitudes}
\begin{figure}[!t]
  	\begin{center}
  		\includegraphics[width=8.5cm]{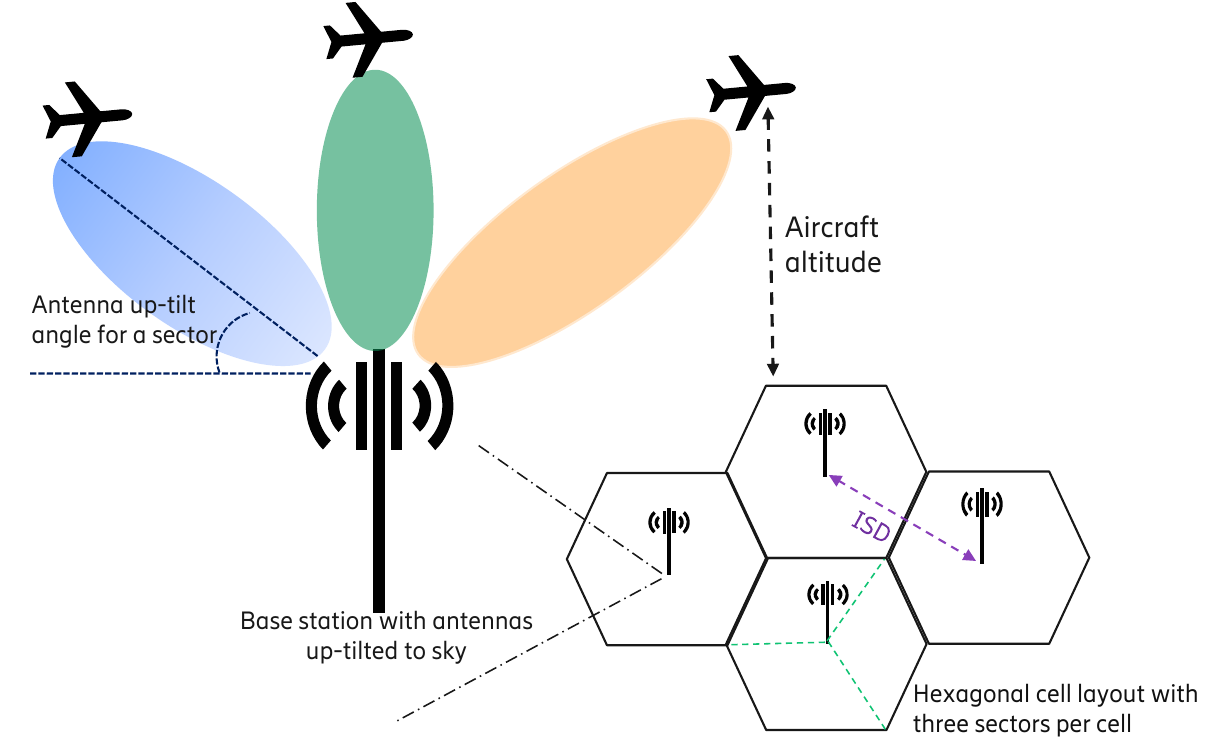}
  		\vspace{-0.02cm}
  		\caption{Illustrative example of a dedicated A2G network. }\vspace{-0.4cm}
  		\label{Fig5}
  	\end{center}
  \end{figure}

In addition to low-altitude UAVs, cellular networks can also provide connectivity to UAVs and/or manned aircraft flying at high altitudes (e.g., a few kilometers). Direct A2G communications using cellular infrastructures can deliver reliable, high data rate, and low latency services to high altitudes \cite{chen20205g}. Remarkably, such cellular-based A2G communication, if properly designed, can offer broadband in-flight connectivity to airplanes that fly at altitudes around 10 km.  To achieve an optimized performance, A2G communications can have dedicated terrestrial BSs similar to cellular towers, yet, with different deployment parameters in terms of e.g., antenna tilts, inter-site distance (ISD), and number of sectors per site (as illustrated in Fig.\,\ref{Fig5}). 

The efficient deployment of A2G networks requires modeling the behavior of complex 5G systems and optimizing a wide range of interdependent parameters. In this regard, ML tools can play an important role in modeling and optimization of dedicated A2G networks in 5G and beyond. To demonstrate this, we look into an example of A2G network deployment optimization using deep learning. Here, we consider the problem of BS's antenna tilts optimization for maximizing the user throughput in the sky. This problem can be tackled using a bi-deep neural network (bi-DNN) architecture proposed in our previous work \cite{chen20205g}. The bi-DNN is composed of two DNNs: the first DNN is developed for approximating the A2G network behavior, and the second DNN is designed as a function optimizer to determine the optimal network settings in terms of various parameters including BS's antenna tilts. The optimal antenna tilts depend on ISD, number of sectors, aircraft's altitude, locations of BSs, and the traffic load. 

In Table \ref{Table1}, we provide a set of representative results for optimal antenna tilt angles for different numbers of sectors and ISD values \cite{chen20205g}. These results are based on the following parameters: 12 km aircraft's altitude, 35 m BS's height, and 3.5 GHz carrier frequency. Also, the performance metric is the 50th percentile user throughput in the sky. One observation is that the optimal antenna tilt angles decrease as the ISD increases to ensure sufficient coverage throughout the cell. For example, for the three-sector case, the optimal tilt angles for different sectors decrease, respectively, from 57, 58, and 57 degrees to 29, 32, and 27 degrees when the ISD increases from 20 km to 80 km. It should be noted that further optimization can be done for the A2G network deployment based on the scenario, system parameters, and requirements. Clearly, an efficient design of dedicated A2G networks is a key step towards enabling a limitless connectivity in the sky.

\begin{table}[t]
	\centering
		\caption{\small Optimized antenna tilts (in degree) for an A2G network for different number of sectors and ISD values.}
		\vspace{-0.25cm}
		\label{Table1}
	\begin{center}
		\begin{tabular}   {|m{1.3cm}|m{2.5cm}|m{2.5cm}|}
			
			\hline \centering
			Number of sectors&\centering	Antenna up-tilt angles for different sectors (ISD=20 km)& \begin{center}	Antenna up-tilt angles for different sectors (ISD=80 km)\end{center} \\ \hline \centering
			1&\centering	78& \hspace{1 cm}	35  \\ \hline  \centering
			3&	\centering$[57, 58, 57]$ &	\hspace{0.6 cm}$[29, 32, 27]$  \\ \hline \centering
		    4& \centering	$[90, 58, 62, 61]$&	\hspace{0.42 cm}$[26, 35, 36, 37]$  \\ \hline
		\end{tabular}\vspace{-0.1cm}
\end{center}
\end{table}

\subsection{On the path to 6G: Complementing terrestrial networks with non-terrestrial networks} 
  Complementing terrestrial cellular networks with NTNs is indispensable towards a connected sky. In particular, complementing terrestrial with satellite links is important beyond the ability to operate in remote areas.  
   It is envisioned that 6G will likely provide integrated solutions with capabilities such as handover or simultaneous use between terrestrial and satellite within a single modem.  An NTN utilizes a space-borne or airborne payload to embark a network node for communication. A space-borne platform refers to a satellite which may be placed into low Earth orbit (LEO), medium Earth orbit (MEO), or geosynchronous Earth orbit (GEO).  An airborne payload refers to a high altitude platform station, which can take the form of airplane, balloon, or airship.
  
  While terrestrial cellular networks have become increasingly widely deployed, there are remote areas where it is cost prohibitive to build terrestrial cellular networks. Meanwhile, it is not uncommon to find UAV use cases in remote areas. One example use case is pipeline inspection and monitoring, in which the inspectors rely on UAVs taking photos and videos of pipelines in remote areas and sending the data back for processing. UAVs also find uses in forestry applications and can assist in wildfire monitoring, forest health monitoring, wildlife tracking, etc. To provide connectivity to the UAVs flying in the remote areas, it is indispensable to complement terrestrial cellular networks with NTNs by utilizing high-altitude platform stations and satellite technology.
  
  UAVs play a key role in mission critical applications such as disaster recovery and public safety missions. Terrestrial cellular based mission critical networks are the main enabler to deliver secure, resilient and quality connectivity to support mission critical services including connecting UAVs. When terrestrial cellular networks are disrupted during disasters (e.g., earthquakes), NTNs are a viable backup alternative to connecting UAVs to support rescue and relief efforts. Thus, complementing terrestrial cellular networks with NTNs is beneficial for UAV connectivity in mission critical applications.
  
  Direct A2G communication can offer large system capacity and short latency and is particularly suitable for providing connectivity to continental flights. However, it is difficult to rely on direct A2G communication for providing connectivity to intercontinental flights over the ocean, which calls for satellite-based inflight connectivity solutions. The European Aviation Network integrates an S-band satellite connection with an LTE-based terrestrial network for providing in-flight connectivity in Europe. An aircraft is equipped with different antennas used for satellite connection and terrestrial connection, respectively. The satellite connection takes over when the aircraft is not covered by the terrestrial network~(e.g.,~over~the~sea). Finally, it is worth noting that 3GPP has been working on evolving NR to support NTN \cite{lin20195g}. In Release-15, 3GPP completed a study on NTN scenarios and channel models. In Release-16, 3GPP completed a further study on solutions to adapting NR to support NTN. In the ongoing Release-17, 3GPP is conducting a work item to enable NR to support NTN.

%

%

{\color{black}	
\section{Conclusions and Future Work}\vspace{-0.01cm}
In this paper, we have shared the learnt lessons about the key barriers and their potential solutions for widespread commercial deployment of flying UAVs in beyond 5G and 6G wireless systems. Furthermore, we have described the capability of cellular networks for offering robust connectivity, reliability, security, and safety of UAVs by relying on their advanced features as well as the UTM support. We have also shared the learnt lessons about how AI can be employed for designing and optimizing cellular-connected UAV networks.

The future of connected sky is promising, despite the challenges we need to overcome on the path to 6G. There are many fruitful avenues for future research. This paper has highlighted two promising areas for future research: (i) offering wireless connectivity to high altitudes with A2G communications, and (ii) the integration of NTNs with terrestrial networks to improve the resilience of UAV connectivity. To pave the way for widespread commercial deployment of flying UAVs, conducting medium- to large-scale experimentation and trials will be another key research direction. Recently, the Perseverance rover carrying a drone helicopter landed on Mars. This may inspire interplanetary UAV communications and networking to become a new research frontier on the path to 6G.}


\def\baselinestretch{1.00}
\bibliographystyle{IEEEtran}
\bibliography{referenceConf}
\vspace{0.5cm}

\textbf{Mohammad Mozaffari} (M'15) is a senior researcher at Ericsson Research, USA. He received the Ph.D. degree in electrical and computer engineering from Virginia Tech. \vspace{0.2cm}  

\textbf{Xingqin Lin} (M'13-SM'20) received the Ph.D. in electrical and computer engineering from The University of Texas at Austin, USA. He is currently a Master Researcher and Standardization Delegate at Ericsson. \vspace{0.2cm} 

\textbf{Stephen Hayes} is the Vice President of Standards and Industry Initiatives for Ericsson Americas. He serves as the Vice-Chair of 3GPP TSG-RAN. 

\end{document}